**Thermally induced chemistry of meteoritic complex organic molecules: a new heat-diffusion model for the atmospheric entry of meteorites**

Christopher N. Shingledecker
225 Colonnade Dr, #8
Charlottesville, VA 22903
cns7ae@VIRGINIA.EDU
434-249-5907
*Department of Chemistry, University of Virginia*

Temperature Gradients in Meteorites

Research over the past four decades has shown a rich variety of complex organic molecular content in some meteorites. This current study is an attempt to gain a better insight into the thermal conditions experienced by these molecules inside meteorites during atmospheric entry. In particular, we wish to understand possible chemical processes that can occur during entry and that might have had an effect on complex organic or prebiotic species that were delivered in this way to the early Earth. A simulation was written in Fortran to model heating by the shock generated during entry and the subsequent thermal diffusion inside the body of a meteorite. Experimental data was used for the thermal parameters of several types of meteorites, including iron-nickel and several classes of chondrites. A Sutton-Graves model of stagnation-point heating was used to calculate peak surface temperatures and an explicit difference formula was used to generate thermal diffusion profiles for both chondrites and iron-nickel type meteorites. Results from the simulation show pyrolytic temperature penetration to a depth of ca. 0.5-1 cm. Non-dissociative warming of meteorite interiors penetrates further to ca. 4 cm. These results support the findings that extraterrestrial delivery is a viable option for prebiotic molecular "seeding" of a planet.

# 1. Introduction

Amino-acids are organic molecules used by living things on Earth in the production of proteins and represent one type of interplanetary complex molecules, i.e. molecules comprised of more than ca. 6 atoms. Their presence in meteorites has been known for quite some time (Kvenvolden *et al.*, 1971). More recently, it was found that dust samples from the comet Tempel 1 as collected by the NASA Stardust mission contain them as well (Glavin *et al.*, 2008). Both meteorites and comets are considered possible sources of early prebiotic molecules that could have been incorporated into the early stages of life on this planet (Basiuk & Douda, 1999).

Though as yet undetected in the interstellar medium, it is thought that amino-acids can form in warm regions such as hot-cores and in protoplanetary disks (Garrod, 2013) via a theorized mechanism involving thermally mobile radical species on dust particles. It is in protoplanetary disks that it is possible for amino-acids to become incorporated into meteorites via, at least initially, the electrostatic attraction and subsequent coagulation of dust-grains on which amino-acids and other complex organic molecules are frozen, though it is unknown whether the pre-biotic molecules form in the protoplanetary disk itself, or earlier, in the interstellar clouds.

Meteoritic samples of amino-acids have been shown to be non-racemic, having a slight L-enantiomeric excess (Engel *et al.*, 1990). The fact that most living things on Earth utilize only L-amino-acids provides support to the extraterrestrial delivery of prebiotics hypothesis mentioned above. Later isotopic analyses provide further support to the claim that the complex organics present were not the result of terrestrial contamination (Engel & Macko, 1997). Questions remain, however, about the differing D/L ratios measured in fragments of the same parent meteorite (Engel & Macko, 2001). Also, if the survival rates of amino-acids in meteors during atmospheric entry are low, then they would be unlikely to have been a major contributor to the in fall of complex organics.

One major challenge to the theory of an extraterrestrial delivery of prebiotic molecules is the well-known effect of atmospheric heating encountered by entering bodies. During the years of the "space race," the phenomenon of atmospheric heating was studied intensively (Ames, 1953; Sutton & Graves, 1971; Tauber, 1989) as it represented a profound danger to the craft and crew of planned future missions. The results of this research showed that peak heating temperatures of over 1000 K can easily be reached. These temperature levels were one factor that called into question whether the complex organics found in meteoritic samples were of extraterrestrial origin or were instead terrestrial contaminants (Engel & Macko, 1997). This is important for the survivability of complex organics since they undergo pyrolysis at high temperatures. Specifically, amino-acids begin to pyrolize at ca. 700 K (Basiuk & Douda, 1999): well below the peak temperatures that can be reached on the whetted surface of a body undergoing atmospheric entry.

There have been several attempts to examine the internal conditions of meteorites during entry. Work by Sears & Mills (1973) and Sears (1975) was particularly important, since theoretical temperature gradients were compared with experimental ones derived from thermoluminescence data for chondrites. There have been other, more advanced simulations that take into account fracturing (Pecina & Ceplecha, 1984; Ceplecha *et al.*, 1993), but it is not the goal of this study to examine the behavior of meteorites with this level of detail.

This paper describes our code, the first version of the Simulation of Extraterrestrial Amino-acid Delivery (SEAD-1), which was written to examine, to a first approximation, the thermal conditions that exist in meteors during atmospheric entry with the specific goal of understanding the internal conditions, particularly temperatures, and how they may affect the chemistry and survivability of complex prebiotic molecules.

## 2. Theoretical Methods

### 2.1 Initial Temperature

The initial temperature of the body in interplanetary space was considered first. Though temperatures in the interstellar medium are typically quite low (ca. 10 K for a dense cloud), temperatures in the vicinity of a star are significantly higher. The internal equilibrium temperature SEAD-1 calculates is based on a standard gray-body formula (Butler, 1966). The fundamental formula for determining the temperature of a body at some distance from a stellar source is given by

$$T = \left( \frac{A_1}{A_2} \frac{a_T}{\sigma \, e_T} \frac{H}{D^2} \right)^{\frac{1}{4}} \qquad (1)$$

where $\sigma$ is the Stefan-Boltzmann constant, $A_1$ is the area of the body that is illuminated by the stellar object, $A_2$ is the total surface area, $H$ is the stellar flux density (the solar constant, in this case), $a_T$ is the absorbance as a function of $T$, $D$ is the distance in AU, and $e_T$ is the emittance, which is also a function of $T$. Following Butler (1966), we assign a value of 0.93 for the absorptance, corresponding to that of a dark chondrite. The emittance was given a value of 0.81, in agreement with experimental results.

As shown in equation (1), it is clear that orbital distances can have a substantial effect on the internal equilibrium temperature. For instance, at 0.4 AU, about a distance from the Sun equal to that of Mercury, a spherical graybody will have an internal temperature of 450 K. On the other hand, at 30 AU, about the distance of Neptune from the Sun, the internal temperature would be 52 K. Distance, therefore, can have an effect on meteoritic complex organic molecules since either close (< 1 AU) or eccentric orbits that take the object close to the Sun may result in heating that could affect complex organic molecular content. The time it takes for a spherical body to reach thermal equilibrium (Carslaw & Jaeger,

1986) is given by

$$t_{eq} \approx \frac{0.5r^2}{Z} \qquad (2)$$

where $Z$ is the thermal diffusivity of the material.

In the model, the temperature of a meteoroid, i.e. a pre-entry meteorite in space, is calculated at runtime using equation (1), based on a previously defined radius, as given by the user. The simulation assumes that the object is at internal thermal equilibrium and a radial distance from the Sun of 1 AU.

## 2.2 Flight Time

The flight time is another critical aspect in considering internal heating during entry. The descent of an object to Earth through the atmosphere can be divided into two periods. The initial descent period characterizes the initial velocity of the meteoroid before entry. These velocities are typically in the range of 11 - 72 km s$^{-1}$. Due to the low density of the upper atmosphere, the entering meteor experiences only moderate drag; however, drag increases below ca. 12 km due to higher atmospheric density. At what is called the retardation point when the drag forces equal the gravitational forces acting on the meteorite, an entering object will reach terminal velocity (between 0.09 and 0.18 km s$^{-1}$) (Hawkins, 1964). Our code utilizes the method described in Irwin (2012) to approximate the flight time of an object following a linear trajectory through the atmosphere. It assumes that most of the initial kinetic energy is dissipated by the drag forces experienced by the object, described by

$$\int_0^{l_D} D \, dx = E_K = \frac{1}{2} m v_0^2 \qquad (3)$$

where $D$ is the drag force, m is the mass of the object, $\upsilon_0$ is the initial entry velocity, and $l_D$ is the distance travelled through the atmosphere. Assuming the average gravitational potential energy is much less than the initial kinetic energy, one can express the average drag force as

$$\overline{D} = \frac{1}{l_D} \int_0^{l_D} D \, dx \; . \qquad (4)$$

Upon integration, this yields

$$\overline{D} l_D = (\overline{q} C_D A) l_D \approx \left( \left[ \frac{1}{2} (\overline{\rho v^2}) \right] C_D A \right) l_D \qquad (5)$$

from which one can calculate the length of flight,

$$l_D \approx \left( \frac{m}{\overline{\rho} C_D A} \right) \left( \frac{\overline{v}}{v_E} \right)^2 \qquad (6)$$

Here, $\overline{\rho}$ is the average air density, $\overline{v}$ is the average velocity, $C_D$ is the drag coefficient, and $A$ is the reference drag area or whetted surface. The drag coefficient was given a value of 0.8, approximating that of a blunt-nosed (in this case, spherical) object, while the average velocity here was approximated to be the terminal velocity of the falling body as calculated using the well-known formula

$$\bar{v} \approx v_{term} = \sqrt{\frac{2mg}{\rho A C_D}} \ . \tag{7}$$

With these values having been calculated, the max time-of-flight is given by

$$t_{flight} \approx \frac{l_D}{\bar{v}} \tag{8}$$

and the related angle of entry, given a straightforward geometric model of descent, is then simply

$$\theta_{entry} \approx \frac{h_{atm}}{l_D} \ . \tag{9}$$

In our code, it is assumed that peak heating is achieved halfway through the total time of flight, i.e. after the density of the atmosphere increases such that the entering object is at terminal velocity.

**2.3 Heat Flux and Peak Surface Temperature**

One of the most important components of calculating the internal temperatures of objects entering the atmosphere is the determination of heat fluxes and peak surface temperatures.

The mechanism for heating during atmospheric entry is complex. The main contribution to heating is not, in fact, due to friction between the surface of the body and atmospheric air molecules. Wind tunnel studies have shown that, at high speeds, a bow shock forms in front of the object, separating the whetted surface from the direct effects of friction (Allen, 1957). Air passing through the bow shock in front of the object is greatly heated, and the formation of a stagnation point (where the velocity of the air is zero) results in convective heat transfer to the whetted surface. Since air temperatures in these regions are so high, the constituent molecules are dissociated; however, after passing through the shockwave, they can recombine exothermically, resulting in another major source of heat that is transferred to the object. Thus, the main contributions to the heating are convection and the heat of molecular recombination (Tauber, 1989).

Sutton and Graves (1971) developed a very convenient general formula for determining the stagnation-point heating for an axisymmetric blunt body in an atmosphere composed of some arbitrary mixture of gases, which is used in SEAD-1 for calculating the heat flux, $\dot{q}$, experienced by the object:

$$\dot{q} = A v_\infty^3 \sqrt{\frac{\rho_\infty}{r}} \tag{10}$$

where here, the Sutton-Graves $A$ value is set to $1.75 \times 10^{-4}$, $v_\infty$ is the free-stream velocity (i.e. the velocity the air would have if the object were stationary), $\rho_\infty$ is the free-stream density, and $r$ is the radius of the blunt object (i.e. the "nose").

Once the stagnation-point heat flux is known, the peak surface temperature can be calculated. Upon entry, the surface temperature of an entering body begins to rise rapidly (Irwin, 2012). At some point (ca. 1000 K), the surface starts emitting heat at a rate proportional to its emissivity ($\varepsilon$). At equilibrium, the surface temperature ceases to rise (i.e. remains constant), as a result of the radiative cooling rate matching the rate of heating. At this point, the temperature

can be expressed as

$$T = \left(\frac{\dot{q}}{\epsilon\sigma}\right)^{\frac{1}{4}}$$ (11)

where $\sigma$ is the Stefan-Boltzmann constant. Our code assumes that after peak heating the surface temperature remains constant.

## 2.4 Temperature Gradients

After the stagnation-point heat flux and surface temperature have been determined, one-dimensional internal temperature gradients are generated. The basis for the temperature gradients is the standard conduction equation

$$k\frac{\partial^2 T}{\partial x^2} + \dot{q} = \rho c \frac{\partial T}{\partial t}$$ (12)

where $c$ is the specific heat of the material, and $\rho$ in this instance is the material density. This is then numerically evaluated using the explicit difference equation (Irwin, 2012)

$$T_{i,m+1} = T_{i,m} + F_O(T_{i+1,m} - 2T_{i,m} + T_{i-1,m}) \, .$$ (13)

In the formula above, $k$ is the thermal conductivity of the material and $F_O$ is the Fourier number, a dimensionless parameter that describes heat conduction:

$$F_O = \alpha \frac{\Delta t}{\Delta x^2} \, .$$ (14)

Here, $\alpha$ is the thermal diffusivity, given by $\alpha = \kappa / \rho c$.

The explicit difference equation, (13), describes heat transfer in a solid body to some depth, $D$, that has been divided into $N$ nodes ($\Delta x$), over a total amount of time which itself is broken up into $M$ time steps ($\Delta t$). For all nodes other than the boundary nodes ($i = 1$ or $N$), the temperature at a given time depends on that node's previous temperature, as well as the previous temperatures of the surrounding nodes ($i - 1$ and $i + 1$). The surface node is maintained at a constant temperature, given by equation (11), and the terminal node ($i = N$) has a temperature of

$$T_{i=N,m+1} = (2F_O)T_{i=N,m} + (1 - 2F_O)T_{i=N-1,m} \;. \tag{15}$$

A requirement for the numerical stability of the results is that the Fourier number, $F_O$, be less than 0.5. Values greater than 0.5 will result in a supercritical pitchfork bifurcation of the solutions.

## 2.5 Parameters

The parameters used in the model (given in Table 1) are from experimental studies by Opeil *et al.* (2010) (hereafter, OEA) utilized a Quantum Design Physical Properties Measurement System. OEA were able to measure the thermal conductivity of six meteorites (both iron-nickel and chondritic). These studies represent the best determination of meteoritic thermal conductivities to date. Previously, values for thermal conductivity were calculated from thermal diffusivity data based on work done by Yomogida and Matsui (1983). Because of the lack of good experimental data on the thermal properties of meteorites and asteroids, values for well-studied materials such as olivine, $(Mg^{+2}$ or $Fe^{+2})_2 SiO_4$, were used in simulations of thermal behavior.

Our simulation uses as input the parameters listed in Table 1. Of note is the temperature behavior of $\kappa$. OEA note that the thermal conductivities for iron-nickel type samples show roughly linear increases with temperature, i.e. $\kappa \propto T$. The chondritic samples, however, show only a slight temperature variation above 100 K.

## 3. Results

Temperature gradients generated by our code are given in Figures 1 and 2 for both carbonaceous chondrite and iron-nickel type meteorites, respectively. A clear result of the calculations is that the depth at which pyrolytic temperatures (ca. 700 K) penetrate is very shallow. Temperature gradients for NWA 5515 show pyrolytic depths reaching only to less than 1 cm, while those for the iron-nickel Campo del Cielo are about twice as large due to the higher thermal diffusivity of the metal. Furthermore, as one would expect, the shape of the gradient depends on the radius of the body. Figure 3 shows a plot of pyrolytic depth (in both mm and as a percentage of the total radius) versus radius using the parameters for NWA 5515. This shows that, in terms of absolute distance, the limit for pyrolytic depths is ca. 5 mm. In addition, as a percentage of radius, the value quickly decays to zero as radius increases. This behavior is again shown in figure 4, which shows the peak surface temperature as a function of radius. From this, one can see that smaller bodies experience greater surface temperatures than larger ones. This possibly counterintuitive result is due to the inverse relationship between area of the whetted surface and peak temperature. Despite the high surface temperatures, however, our results indicate that the majority of the interior remains unheated.

One can also see in FIG 1 & 2 that three temperature regimes can be broadly categorized in entering meteorites with respect to potential chemical effects. The first is represented by the outermost layers which are heated most strongly. In these pyrolysis regions, one does not expect large complex organics such as amino-acids to survive.  The second type of region is represented by the cold central parts of entering meteorites where the temperatures remain effectively unchanged from their initial values. In this second class of environment, complex organics would likely remain effectively unchanged structurally. The last is represented by the intermediate regions that are heated, though not sufficiently to cause dissociation of the complex organics.

## 4. Discussion

From these data, a picture emerges of the ideal candidates for the extraterrestrial delivery of amino-acids, i.e. a large ($r > 2$m) chondritic meteorite. Such an object will be able to serve as an effective vehicle for complex organic molecules because of the low temperatures that exist in the bulk of the structure. This implies one reason why meteorites represent an important potential delivery mechanism for extraterrestrial pre-biotic molecules, since molecules in the cold regions would survive at least the entry period mostly intact. As previously mentioned, analysis of the chondritic meteorites such as the Murchison show that at least some non-trivial fraction of the original number of complex organic molecules are also able to survive entry and the impact event.

Fragmentation represents a complication to the survival of meteorites. In a fragmentation event, it is possible that break away sections containing more of the whetted surface (or more intensely heated side) will have less organic content than sections which contain the cooler, previously shielded areas. One result of the fragmentation of an unevenly heated parent body may be the racemization of the chiral chemical species. The regions of entering meteors that are of an intermediate temperature, i.e. those that are not sufficiently hot to pyrolize the amino-acids but not cold enough to allow them to remain completely unchanged, could provide enough thermal energy to facilitate racemization of previously non-racemic chiral species (Engel, 2009). Thus, there seem to be two main mechanisms at play which result in the differing D/L ratios observed in fragments of the same parent body. First, there is the intermediate heating prior to fragmentation, with sections closer to, or containing more of, the original whetted surface likely to have more racemic D/L ratios. After fragmentation, one expects the smaller objects to heat more quickly, resulting in further differences between the original interplanetary D/L values and those from an analysis of the fallen body.

The basic results given by the simulation here are also broadly in agreement with previous work (Sears & Mills, 1973; Sears 1975) on the depth of pyrolytic temperature penetration. One simplifying feature of our model is the lack of an ablative term. Ablation, the vaporization of surface layers of a solid due to high temperatures, is a well-studied phenomenon to which objects entering an atmosphere are subject. This phenomenon is closely related to the luminosity seen in meteors, as the surface material gives off light as it ablates (Ceplecha *et al.*, 1993).

Our model represents a simulation of the so-called zero-ablation case of meteoritic heating, in which steady-state conditions are not met (i.e. $\partial T/\partial t \neq 0$), and thus, internal gradients are determined mainly by the duration of heating. This approximation most closely models the intermediate times of atmospheric entry, after the density of the atmosphere increases such that peak-heating is reached, but before the most extreme structural changes that occur at later times closer to the actual impact event. There have been simulations that have treated these physical processes, as well as the complex phenomenon of fragmentation, in great detail (Pecina & Ceplecha, 1984; Ceplecha *et al.*, 1993). Factors that influence the rates of ablation and fragmentation include the density and porosity of the material as well as the unique structural condition of each object (Consolmagno *et al.*, 2008). These detailed physical models, though, give less insight into the conditions inside rocky bodies during atmospheric entry, and specifically how such environments might affect the chemistry of the organic molecular content inside. Thus, SEAD-1 represents the first, and most approximate, version of a simulation designed to look at meteoritic pre-biotic (i.e. organic) chemistry.

## 4. Conclusions

Internal temperature gradients of meteors during atmospheric entry were calculated with the goal of gaining insight into the survivability of prebiotic molecules, specifically, amino-acids. A simulation, SEAD-1, was written to generate these, using the Sutton-Graves formula for stagnation-point heating of an axisymmetric object (equation 10) to calculate the heat-flux during entry and the explicit difference formula (equation 13). Physical thermal parameters came from experimental studies by OEA on 6 different meteorites. Results showed that pyrolytic temperatures (ca. 700 K) penetrate only to ca. 0.5-1 cm for a carbonaceous chondrite, and that, as a percentage of the radius, this value quickly approaches zero as radii increase.

This study indicates that a large fraction of the prebiotic molecular content of a meteoroid could potentially survive intact during entry. Also, differential heating of a body and subsequent fracturing may explain the observed differences in D/L ratios seen in fragments of the same parent body (Engel & Macko, 2001; Engel, 2009). Chiral organic molecules in fragments closer to the whetted surface would be expected to have undergone higher rates of thermal racemization, as would smaller fragments.

In terms of viability as a mechanism for delivery of pre-biotic molecules, meteorite impacts seem to be fairly negligible today. Recent work has shown that the flux of small (1-10 m diameter) objects striking the Earth has a power law distribution with ca. 100 impacts per year for objects with diameters of ca. 1 m. However, it is critical to keep in mind the conditions of the early Earth when extraterrestrial organic molecules may have been most important. Life on this planet is thought to have originated in the Archean eon, 4-2.5 Ga. As shown by the surface of the moon (and indeed by its existence), impacts were a much more frequent occurrence early in the history of the solar system. Thus, the influx of organic content would have been much higher. Comets represent another class of potential vehicle for the delivery of extraterrestrial pre- biotics. In this case, though, cometary impacts were (then as now) likely much less frequent than meteoritic impact. Therefore, in terms of a reliable, regular source of complex molecules, meteorites may have been more important.

The construction of ALMA, the Atacama Large Millimeter Array, marks what may be a turning point in human understanding of the chemical nature of the universe. With it, astronomers will be able to obtain highly resolved maps of environments such as protoplanetary disks. Such images may give a glimpse into the molecular composition of planet forming regions and the chemical resources that could play an important role in the formation of life. Thus, understanding potential delivery mechanisms is a natural extension of our increased knowledge of the astrochemistry of these environments.

## Acknowledgments

I would like to thank Mike Engel, Stephen Macko, and my advisor, Eric Herbst for their kind advice, suggestions, and encouragement.

## Author Disclosure Statement

No competing financial interests exist.

**Tables**

**Table 1**: Experimental values of physical and thermal parameters for meteorites from Opeil *et al.* (2010)

| Meteorite | Density (g cm⁻³) | $k$ (W m⁻¹ K⁻¹) | C (J kg⁻¹ K⁻¹) |
|---|---|---|---|
| Abee (E4) | 3.279 | 5.35 | 500 |
| Campo del Cielo (IAB) | 7.71 | 22.4 | 375 |
| Cold Bokkveld (CM2) | 1.662 | 0.5 | 500 |
| Cronstad (H5) | 3.15 | 1.88 | 550 |
| Lumpkin (L6) | 2.927 | 1.47 | 570 |
| NWA 5515 (CK4) | 2.675 | 1.48 | 500 |

**Figure Legends**

**FIG.1.** Temperature gradients for the carbonaceous chondrite NWA 5515, calculated for a collection of radii.

**FIG.2.** The Campo del Cielo iron-nickel meteorite temperature gradients.

**FIG.3.** Results showing the pyrolytic temperature depth in both mm and as a percentage of the radius with parameters for the NWA 5515 chondrite.

**FIG.4.** Peak surface temperature as a function of radius for the NWA 5515 chondrite.